\documentclass[prl,floatfix,twocolumn,showpacs,amsmath,amssymb]{revtex4}

\usepackage{graphicx}
\usepackage{dcolumn}
\usepackage{bm}

\begin{document}
\title{
Quantum Hall effect and  the topological number in graphene
}
\author{Yasumasa Hasegawa$^1$ and Mahito Kohmoto$^2$}
\affiliation{
$^1$~ Department of Material Science,
Graduate School of Material Science,\\
University of Hyogo\\
 Ako, Hyogo 678-1297, Japan
\\
$^2$~Institute for Solid State Physics, University of Tokyo,
Kashiwanoha 5-1-5, Kashiwa-shi, Chiba, 277-8581, Japan}

\date{8 May, 2006}


\begin{abstract}

Recently unusual integer quantum Hall effect  was observed in  graphene in which
the Hall conductivity is  quantized as  
$\sigma_{xy}=(\pm 2, \pm 6, \pm 10, \cdots) \times \frac{e^2}{h}$, 
where $e$ is the electron charge and $h$ is the Planck constant.
To explain this we consider  the  energy structure as a function of magnetic field 
 (the Hofstadter butterfly diagram) on the honeycomb lattice
and the Streda formula for Hall conductivity.
The quantized Hall conductivity is obtained to be  odd integer, $\pm1, \pm3, \pm5, \cdots$ times two 
(spin degrees of freedom) when a uniform magnetic field is as high as 30T for example. 
When the system is anisotropic and  
described by the generalized honeycomb lattice, 
Hall conductivity can be quantized to be any integer number.
We also compare the results with those for  the square lattice under extremely strong magnetic field.
\end{abstract}

\maketitle



Quantum Hall effect 
is one of the most spectacular phenomena in condensed matter physics.
The integer quantum Hall effect has been observed in two-dimensional electrons in
 semiconductor heterostructure\cite{Klitzing1980,Laughlin1981} 
and in quasi-one-dimensional organic conductors in magnetic-field-induced
spin density wave state\cite{Cooper1989,Hannahs1989,Poilblanc1987,Kohmoto 1990,Machida1994}.
Recently, quantized Hall effect is observed
 in graphene\cite{Novoselov2005,Zheng2005}. 
Unlike the previously studied 
 systems, graphene has a unique property that there exist zero modes  (so-called massless Dirac fermions) 
in the absence of a magnetic field
  at half-filling due to the
honeycomb lattice structure.
Hall conductivity in graphene has been observed to be 
\begin{equation}
\sigma_{xy} = 2 (2n + 1) \frac{e^2}{h} ,
\label{qheinhoneycomb}
\end{equation}
where  $n$ is an integer.
Gusynin and Sharapov\cite{Gusynin2005} have
 explained these unusual quantized Hall effect as a result of four times the
quantum Hall conductivity $(n+1/2) e^2/h$ for each Dirac fermions, where the factor four
comes from the spin degrees of freedom times number of the Dirac fermions in the Brillouin zone.
Although their explanation gives the correct quantum number, their argument may be justified only 
in the low magnetic field limit and the
logic  of quantum Hall conductivity using
Dirac fermions is not correct in general\cite{Oshikawa1994}.
Peres et al.\cite{Peres2006} obtained the same result by solving tight-binding electrons on a 
finite size of the  honeycomb lattice in a magnetic field numerically.

In this letter we give an approach to the unusual quantum Hall effect in graphene
 in terms of topological stability. Zero modes at $H=0$ is a consequence  of the honeycomb lattice structure which was at best implicit in the Dirac Fermion approach. When periodic structure is taken into account, there exist gaps in the presence of a magnetic field. 
When chemical potential is in one of the gaps, 
the Hall conductivity is quantized to be an integer, which 
is a topological number and can be obtained by 
Diophantine equation\cite{TKNN1982,Kohmoto1985,Kohmoto1989}. The  Hall conductivity is also obtained by
the Streda formula\cite{Streda1982}
\begin{equation}
  \sigma_{xy}=\frac{\partial N}{\partial B} \frac{e^2}{h},
\label{streda}
\end{equation}
where $N$ is the total density of states below the gap. 
Note that  Hall conductances  obtained by the Streda formula and by topological numbers are the same since both start with the linear response theory.
Applying the  Streda formula in  the generalized honeycomb lattices in a magnetic field,
we obtain the unusual Hall effect. This unusual Hall effect  is also seen in the square lattice 
at half-filling with a magnetic field close to half flux quantum in plaquette\cite{HK}. 

%
%
\begin{figure}[tb]
\includegraphics[width=0.3\textwidth]{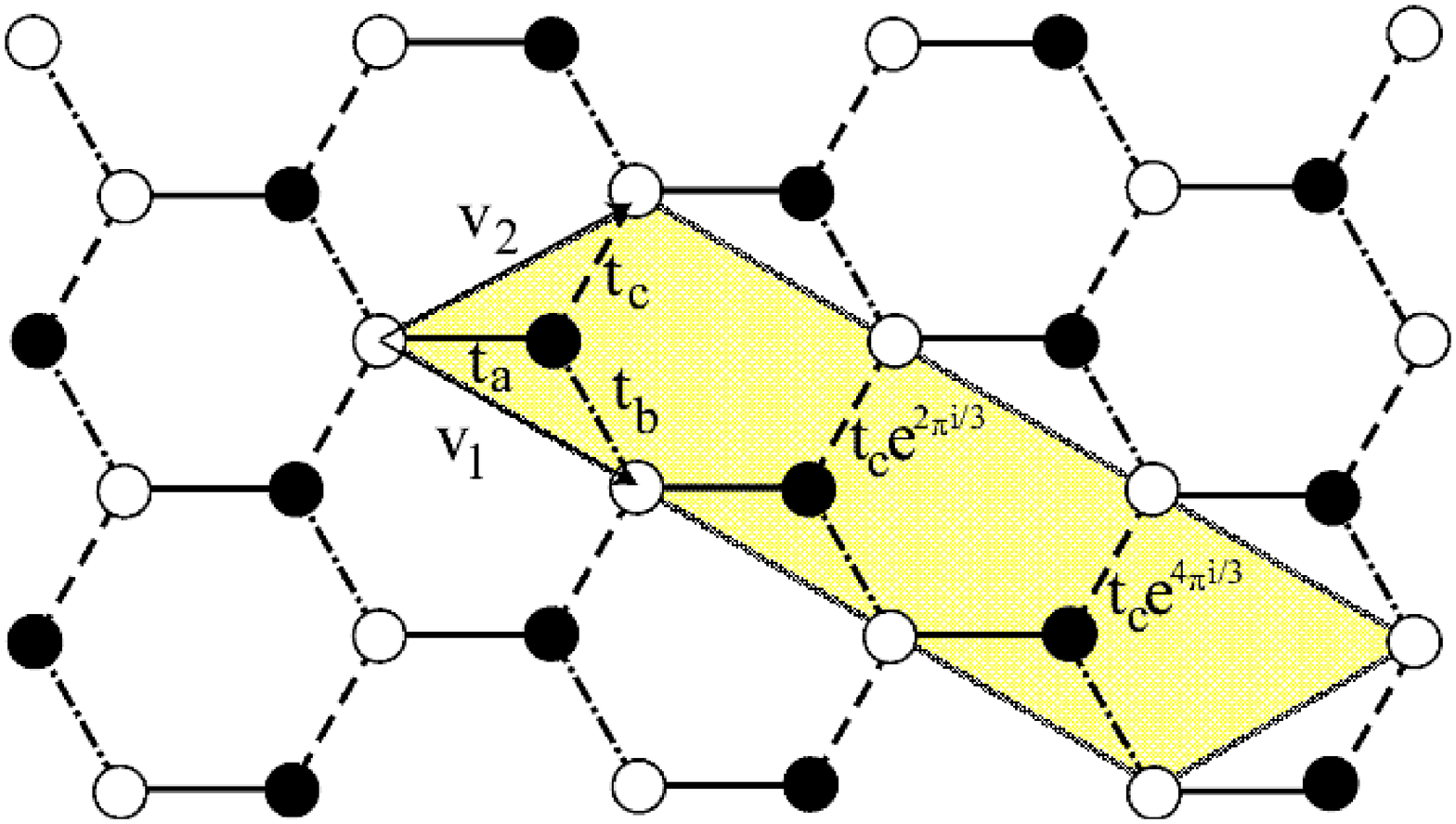}
\caption{Generalized honeycomb lattice. Unit vectors are $\mathbf{v}_1=(\frac{\sqrt{3}}{2}a, -\frac{1}{2}a)$
and $\mathbf{v}_2=(\frac{\sqrt{3}}{2}a, \frac{1}{2}a$),
where $a$ is the lattice constant ($a=0.25$nm).
A yellow parallelogram is a supercell in the presence of a uniform magnetic field of
$\phi=\frac{1}{3}\phi_0$ per unit cell.}
\label{fig1}
\end{figure}
\begin{figure}[tb]
\includegraphics[width=0.3\textwidth]{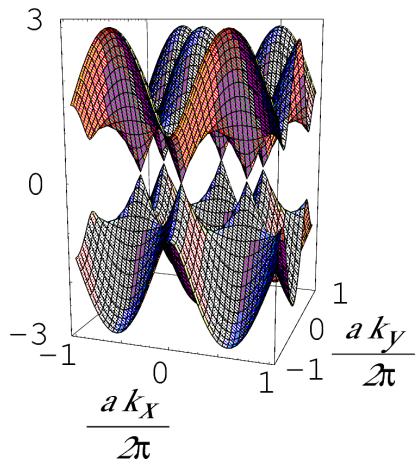}
\caption{3D plot of the energy of the honeycomb lattice.
Zero modes exist at $\mathbf{k}=\pm(\frac{2\pi}{\sqrt{3}a}, \pm \frac{2\pi}{3a})$ and 
$\mathbf{k}=\pm(0,\frac{4\pi}{3a})$}
\label{fig3d}
\end{figure}

%
The unit cell of the generalized honeycomb lattice contains two sublattices as shown 
by the white circles and black circles
in Fig. \ref{fig1}.
There are three non-equivalent hopping matrix elements, $t_a$, $t_b$ and $t_c$.
In the usual honeycomb lattice $t_a=t_b=t_c$, but they may be different for each other under
the uniaxial pressure. 
In the absence of the magnetic field zero modes
exists, if the condition 
\begin{equation}
\left| \left| \frac{t_b}{t_a}\right|-1 \right| < \left| \frac{t_c}{t_a}\right| 
  < \left| \left|\frac{t_b}{t_a}\right| +1\right|
\end{equation}  
is fulfilled\cite{Hasegawa2006}. The energy for $t_a=t_b=t_c$ is plotted in Fig~\ref{fig3d}.

\begin{figure}[tbh]
\includegraphics[width=0.3\textwidth]{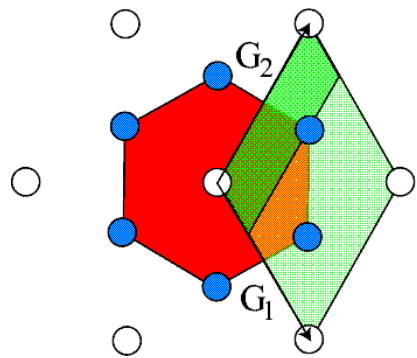}
\caption{Brillouin zone for the honeycomb lattice. 
$\mathbf{G}_1=(\frac{2\pi}{\sqrt{3}a}, -\frac{2\pi}{a})$
and $\mathbf{G}_2=(\frac{2\pi}{\sqrt{3}a}, \frac{2\pi}{a})$.
White circles are $\Gamma$ points. A red hexagon is a first Brillouin zone. 
A Brillouin zone can be taken as the light green diamond.
The zero modes occur at the corner of the
first Brillouin zone (Blue circles) in the absence of a magnetic field.
A dark green parallelogram is a magnetic Brillouin zone in the presence of a uniform magnetic field of
$\phi=\frac{1}{3}\phi_0$ per unit cell.}
\label{fig2}
\end{figure}

We study a system in a uniform magnetic field with a magnetic flux $\phi=\frac{p}{q} \phi_0$ 
per unit hexagon, where $\phi_0=2 \pi \hbar/e = 4.14 \times 10^{15}$~Tm$^2$ is the flux quantum.  The 
magnetic field is $B=\phi/S$, where $S=\sqrt{3}a^2/2$
is the area of unit cell of the honeycomb lattice.
By choosing a suitable gauge and a local gauge transformation,
the effect of the magnetic field can be treated as a phase factor in one of the hopping matrix elements, 
 $t_c$ for example, as shown in Fig.~\ref{fig1}.
We  take $q$ times larger super cell and 
magnetic Brillouin zone is $\frac{1}{q}$ times smaller in the presence of a magnetic field as shown 
in Figs.~\ref{fig1} and \ref{fig2}.  
\begin{figure}[tbh]
\includegraphics[width=0.4\textwidth]{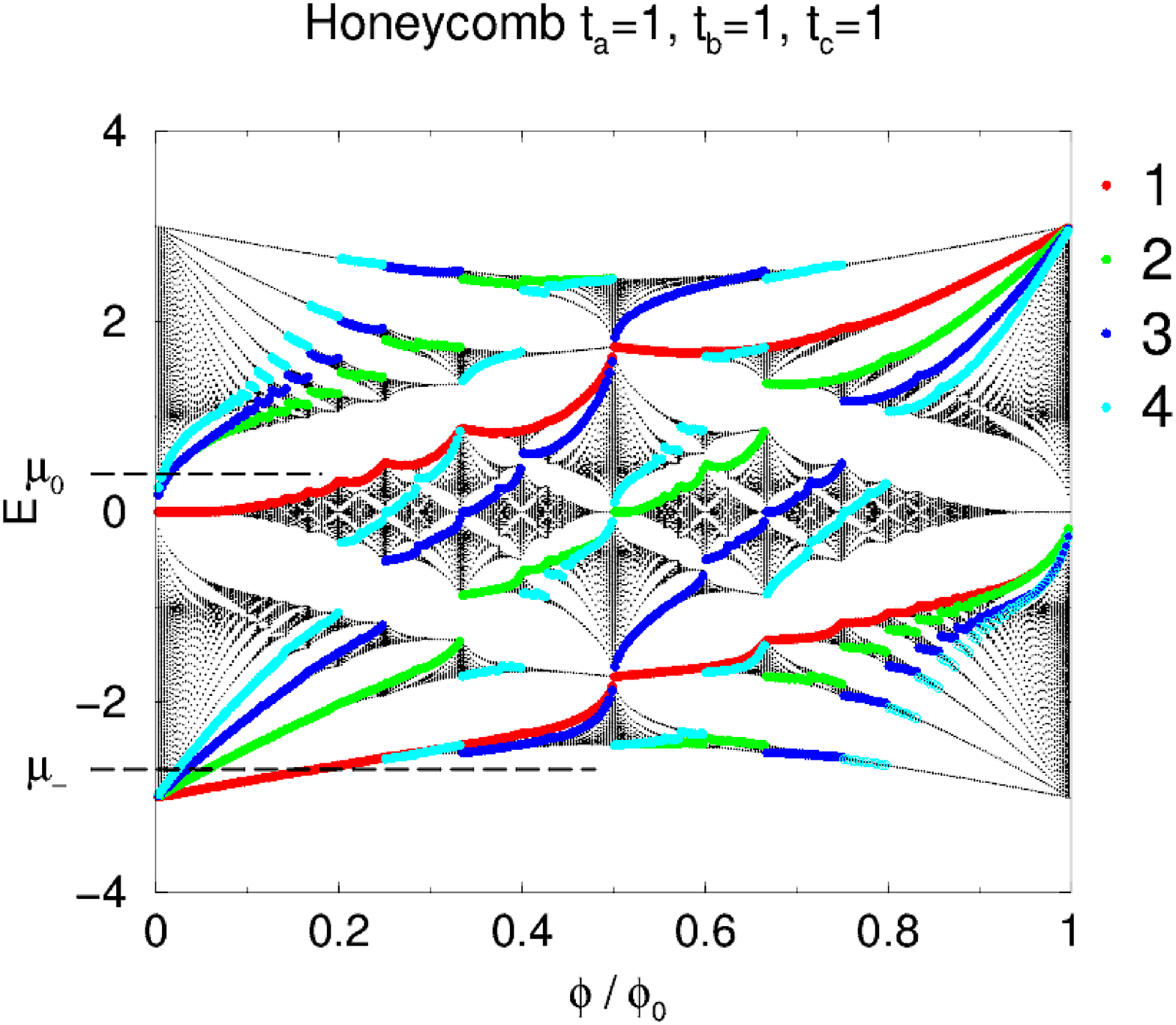}
\caption{Energy spectrum with a magnetic flux $\phi$ in the honeycomb lattice.}
\label{fighoney}
\end{figure}
%
%
\begin{figure}[tbh]
\includegraphics[width=0.4\textwidth]{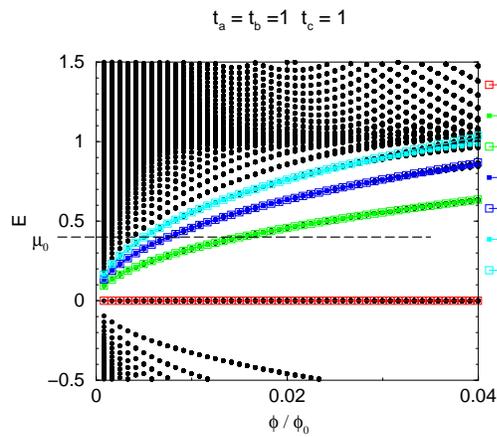}
\caption{The same as Fig.~\ref{fighoney} for $\phi \approx 0$ and $E \approx 0$.}
\label{fighoneyb}
\end{figure}
%
\begin{figure}[tbh]
\includegraphics[width=0.4\textwidth]{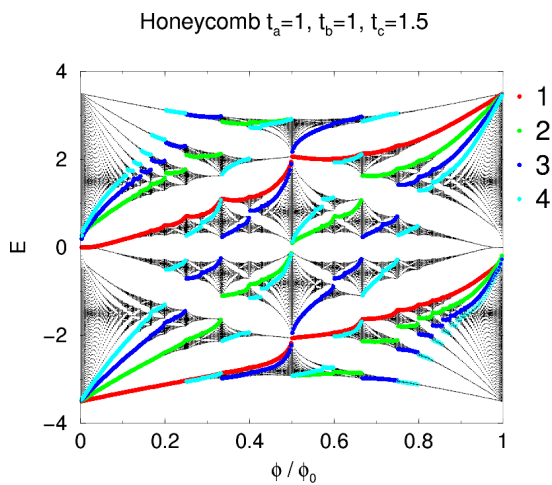}
\caption{Energy spectrum with a magnetic flux $\phi$  in the 
generalized honeycomb lattice with $t_a=t_b=1, t_c=1.5$}
\label{fighoney15}
\end{figure}
\begin{figure}[tbh]
\includegraphics[width=0.4\textwidth]{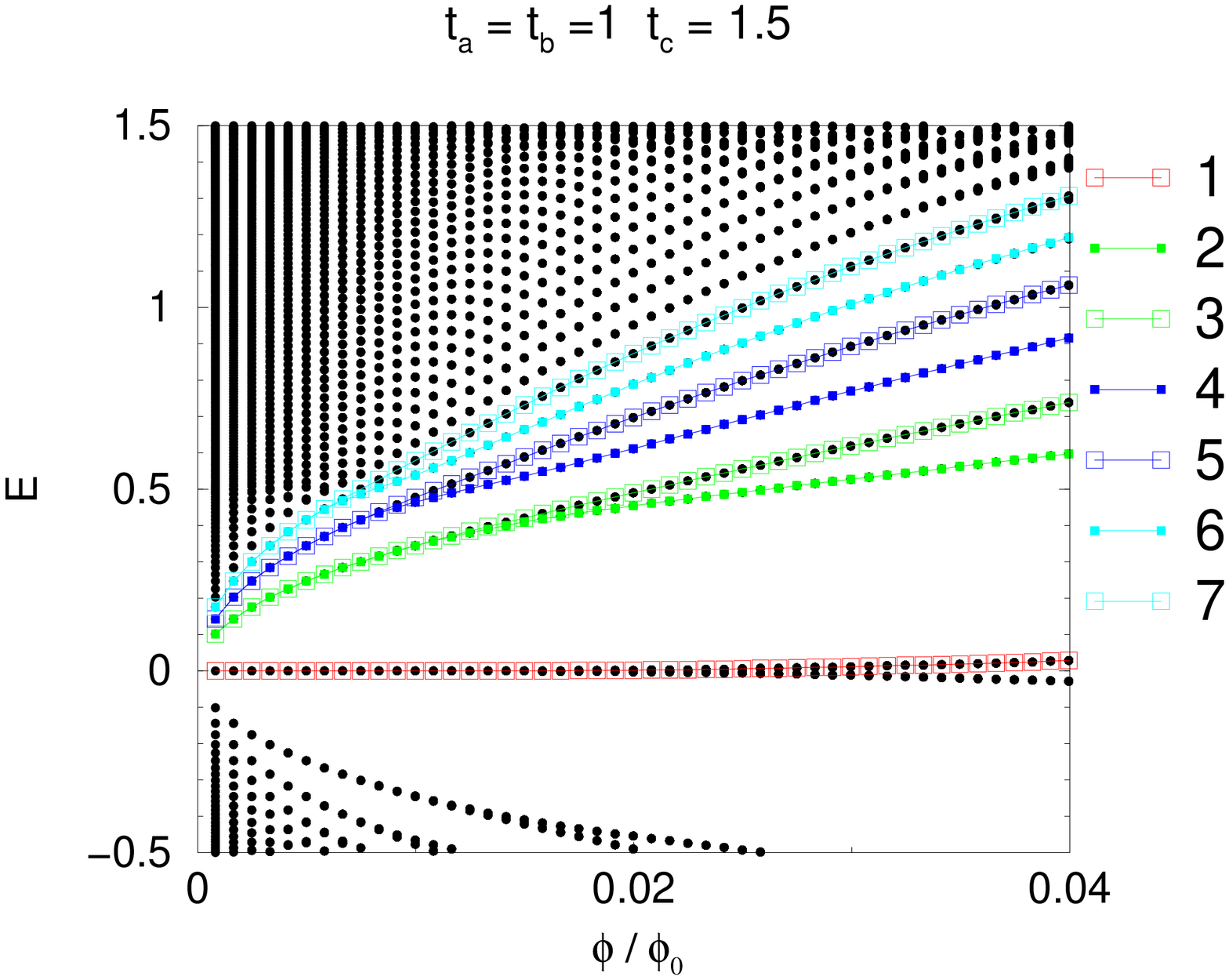}
\caption{The same as Fig.~\ref{fighoney15} in $\phi \approx 0$ and $E \approx 0$.}
\label{fighoney15b}
\end{figure}
%
\begin{figure}[tbh]
\includegraphics[width=0.4\textwidth]{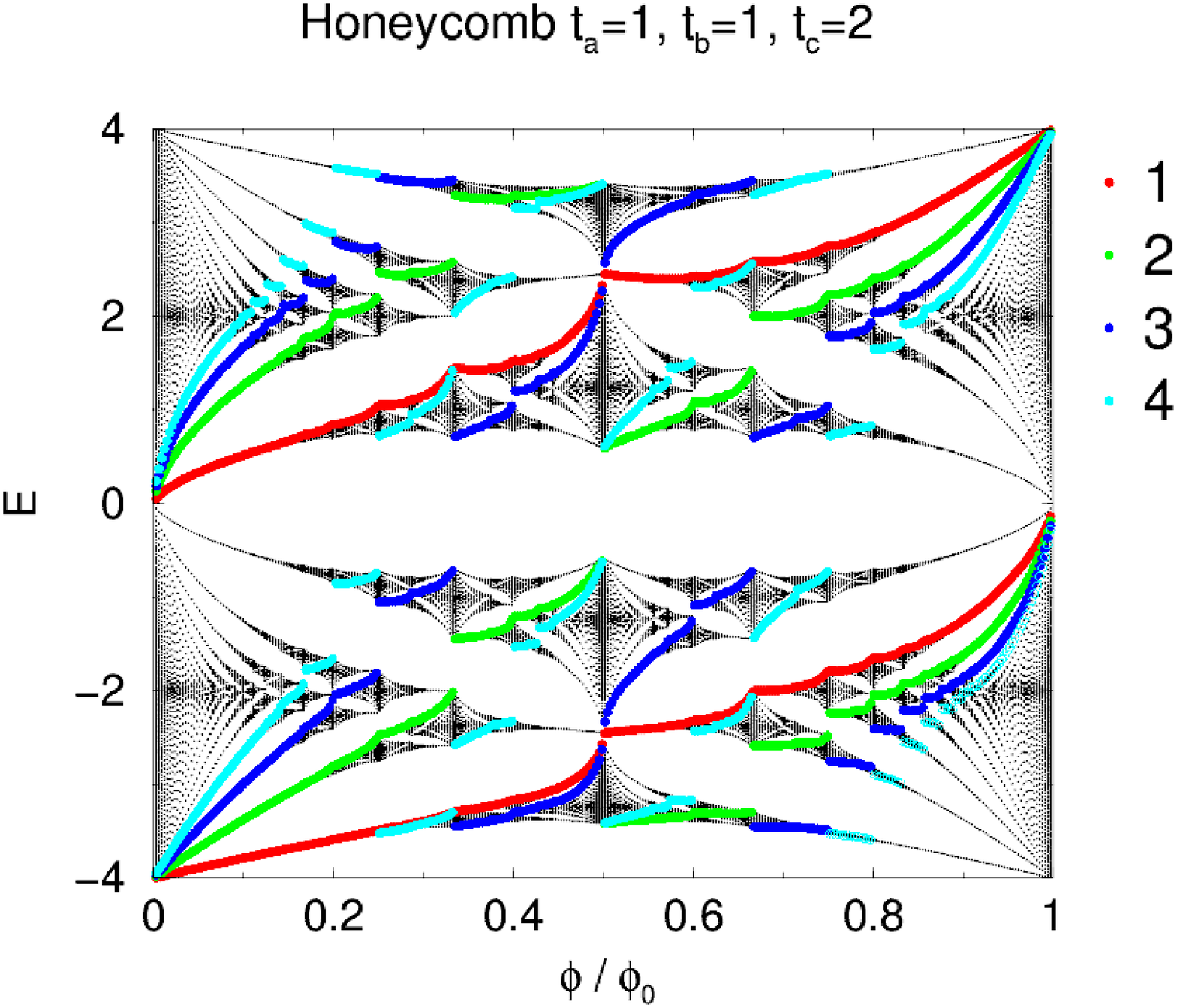}
\caption{Energy spectrum with a magnetic flux $\phi$  in the honeycomb lattice of $t_a=t_b=1, t_c=2$}
\label{fighoney2}
\end{figure}
\begin{figure}[tbh]
\includegraphics[width=0.4\textwidth]{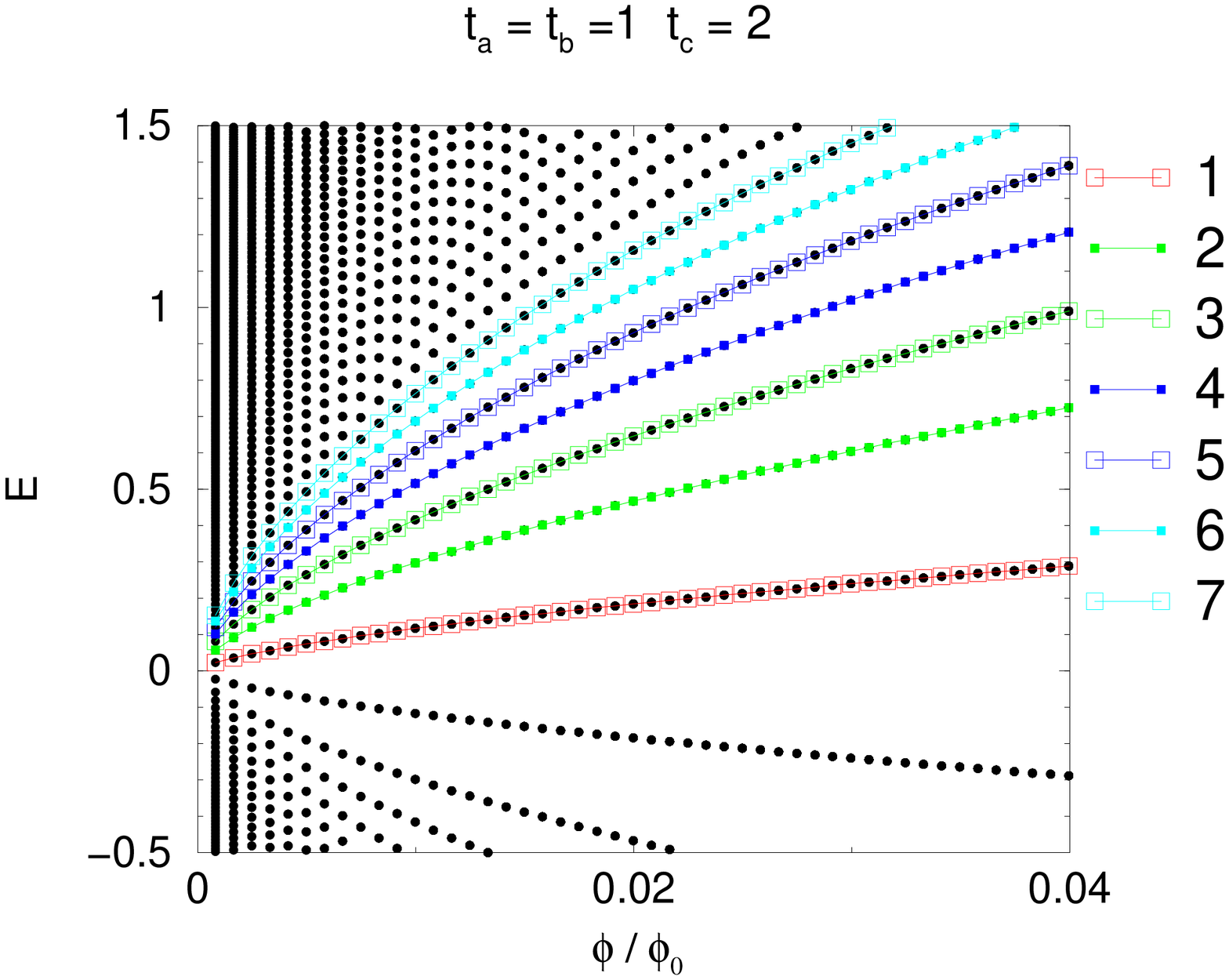}
\caption{The same as Fig.~\ref{fighoney2} in $\phi \approx 0$ and $E \approx 0$.}
\label{fighoney2b}
\end{figure}
\begin{figure}[tbh]
\includegraphics[width=0.4\textwidth]{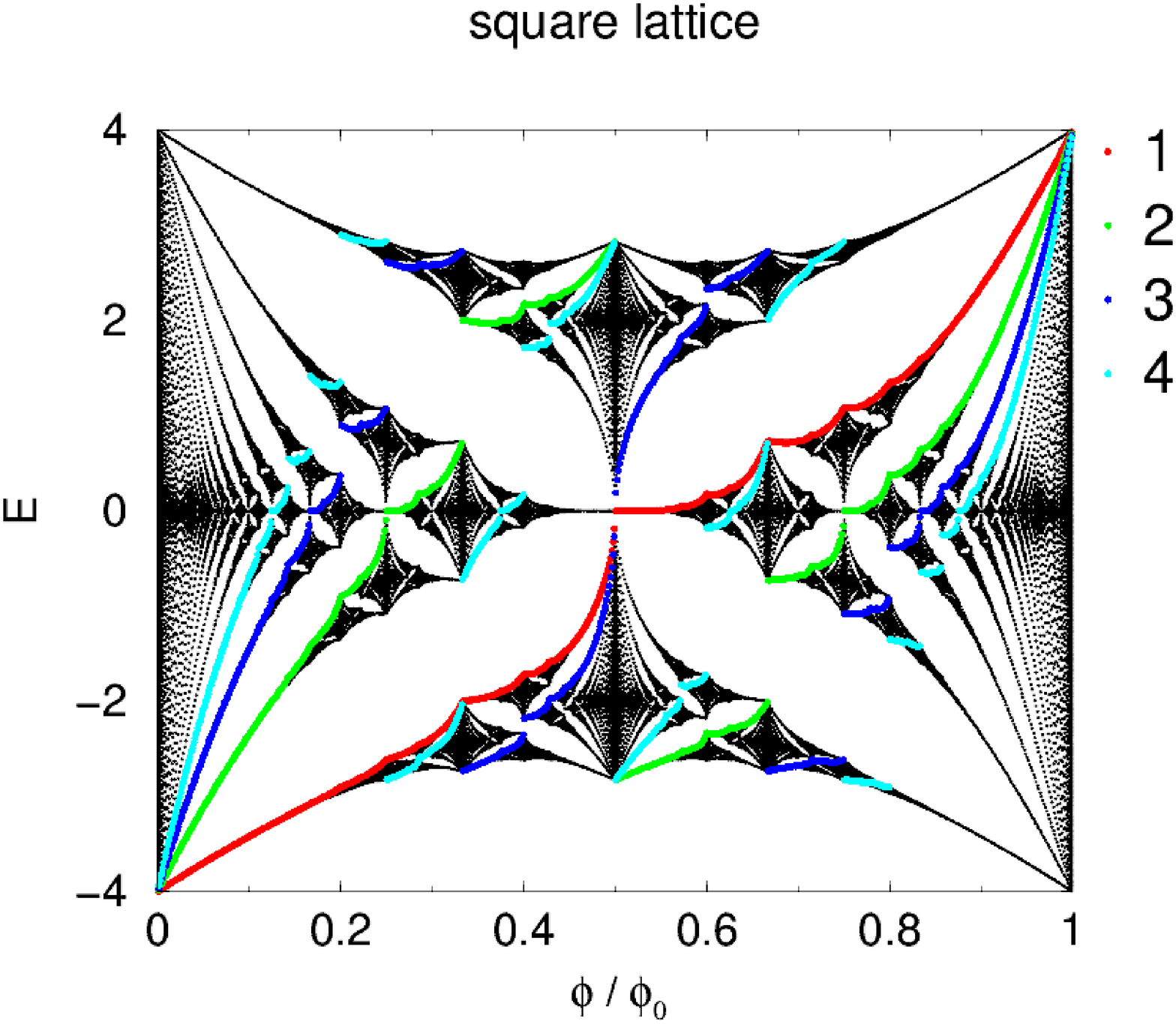}
\caption{Energy spectrum with a magnetic flux $\phi$  in the square lattice. Hofstadter butterfly}
\label{figsquare}
\end{figure}
The energy is obtained numerically by getting the eigenvalues of $2q \times 2q$ matrix 
for each wave vector $\mathbf{k}$ in the magnetic Brillouin zone.
We take the sufficiently large values of $q$ ($q=331$ in Figs.~\ref{fighoney}, \ref{fighoney15} and \ref{fighoney2},
$q=1201$ in  Figs.~\ref{fighoneyb}, \ref{fighoney15b} and \ref{fighoney2b}, 
and $q=701$ in Fig.~\ref{figsquare}), and then
the width of each band is very narrow. In this case we do not have to vary $\mathbf{k}$ to get the energy spectrum
as a function of the magnetic field. 
Then  taking $p=1,2, \cdots, q$, 
 we get so-called Hofstadter butterfly diagram\cite{Hofstadter1976} for generalized honeycomb 
lattice and square lattice\cite{Hasegawa1990,Iye2004}. 
In these figures we also plot the marks which shows the $p$-th, $2p$-th, $3p$-th, $\ldots$,
$n p$-th
energy bands from the bottom in each $\phi=\frac{p}{q}$. We also mark the bands of 
$(n p + m q)$-th from the bottom,
where $m$ is positive or negative integer. 
If there is a gap above the  $(n p + m q)$-th band and the chemical potential is in that gap,
 the Hall conductivity is obtained as
\begin{equation}
 \sigma_{xy}=n \frac{e^2}{h}
\end{equation}
from the Streda formula Eq. (\ref{streda}).
It is seen that there are gaps corresponding to the Landau levels 
 near the bottom of the energy at $\phi \approx 0$ in the generalized honeycomb lattices
(Figs.\ref{fighoney}, \ref{fighoney15} and  \ref{fighoney2}) and square lattice(Fig.~\ref{figsquare}).
Although each Landau level near the bottom of the energy is broadened and splitting into $p$ subband
due to the periodicity of the system in general, the broadening and the mini gaps can be neglected in the 
weak magnetic field limit. 
Therefore, if the chemical potential is fixed near the bottom of the band 
($\mu_-$ in Fig.~\ref{fighoney}),
usual quantum Hall effect $\sigma_{xy}=2 n e^2/h$ with $n= \ldots, 3, 2, 1, 0$ 
is observed as the magnetic field is increased, 
which can be seen in the lower left part 
in Figs.~\ref{fighoney}, \ref{fighoney15} and \ref{fighoney2}.
Here we have assumed the spin degeneracy and multiply a factor of 2.

Interesting phenomena are seen near half-filling ($E \approx 0$).
For the usual honeycomb lattice, visible gaps open only for the odd number $n$ (Fig.~\ref{fighoneyb}). 
The band for $n=2$ and $3$ are overlapped in the scale of the figures. 
Although there are gaps between every 
$2q$ bands when $\phi=\frac{p}{q} \phi_0$ in general,  only 
gaps above $(n p + q)$-th band with odd integer $n$  are
seen  in the  $E \approx 0$ region in that scale.
Therefore, if the chemical potential is fixed near $E=0$
(for example $\mu_0$ in Figs.~\ref{fighoney} and \ref{fighoneyb}),
quantum Hall effect with $\sigma_{xy}= 2 n e^2/h$ with $n= \cdots, 7, 5, 3, 1$, as 
the magnetic field is increased.
If we change parameter $t_c$ in the generalized honeycomb lattice, gaps above the even quantum number
are also visible
 at finite $\phi$. 
The gap between $n=2$ and $n=3$ begins to open at 
$\phi\approx 0.018$ for $t_a=t_b=1$, $t_c=1.5$ (Fig.~\ref{fighoney15b}).
The gap between $n=1$ and $n=0$ becomes open for $\phi \gtrsim 0.03$ for $t_a=t_b=1$, $t_c=1.5$.
If the chemical potential is in that gap, Hall conductivity is zero. 
In the graphene case, $B=400$T corresponds to $\phi \approx 0.005 \phi_0$

When $t_a=t_b=1$ and $t_c=2$, two zero modes in the Brillouin zone merge into 
a  confluent mode at $H=0$\cite{Hasegawa2006}.
In that case gaps are visible at every integer number 
$n$ at $E \approx 0$ and $\phi \approx 0$ (Fig.~\ref{fighoney2b}).

Similar situation as in the honeycomb lattice with $t_a=t_b=t_c$ in a small magnetic field 
 is seen in the square lattice with half flux per unit cell
($\phi/\phi_0=1/2$).
Figure~\ref{figsquare} is the Hofstadter butterfly diagram for the square lattice. 
Note the similarity between $E \approx 0$, $\phi \approx 0.5$ in Fig.~\ref{figsquare} and 
$E \approx 0$, $\phi / \phi_0 \approx 0$ in Fig~\ref{fighoney}.
In both cases visible gaps open only for odd quantum numbers ($n=1,3,5, \ldots$).
For the square lattice with just half flux quantum per unit cell, the supercell is twice as large as the 
square unit cell and there exist two zero modes in the magnetic Brillouin zone.

The quantum Hall effect in the square lattice with nearly half flux per unit cell and in half-filling case 
can be discussed as follows.
When the flux $\phi=\frac{p}{q} \phi_0$ is applied in the square lattice, there are $q$ bands and 
the contribution of the $r$-th
band to the Hall conductivity is given as
\begin{equation}
 \sigma_{xy}^{(r)}=(t_r - t_{r-1}) \frac{e^2}{h},
\end{equation}
where $t_r$ and $t_{r-1}$ are integers solutions of Diophantine equation\cite{Kohmoto1989}
\begin{equation}
  r = s_r q + t_r p,
\label{Diophantine}
\end{equation}
and $t_0 = 0$.
When $r$ bands from the bottom are filled, we get
\begin{equation}
 \sigma_{xy}= \sum_{i=1}^r \sigma_{xy}^{(i)} = t_r  \frac{e^2}{h}.
\end{equation}
Consider the case that electrons are near half-filling,
\begin{equation}
 \frac{r}{q} = \frac{1}{2} + \delta ,
\end{equation}
where $|\delta| \ll 1$ and 
\begin{equation}
 \frac{\phi}{\phi_0} = \frac{p}{q} = \frac{p_0}{q_0} + \eta ,
\end{equation}
where $|\eta | \ll 1$.
From Eq.~(\ref{Diophantine}) we get
\begin{align}
 t_r &= \frac{q_0 (1-2 s_r)}{2p_0} ,
\label{Diophantine2} \\
 \delta &= t_r \eta .
\label{Dio3}
\end{align}
Then we can conclude from Eq.~(\ref{Diophantine2}) that
$q_0$ should be even number ($q_0=2 m$ where $m$ is integer). 
For $q_0=2m$ there are zero modes at $E=0$ \cite{Kohmoto1989}. 
When Eq.~(\ref{Dio3}) is satisfied, the $r$ bands are completely filled
and the Hall conductivity for nearly-half-filled
electrons are quantized as $m$ times odd integers, because $(1-2s_r)/p_0$ cannot be an even number.
This result is the same as the quantum Hall effect on honeycomb lattice in small magnetic field.

In conclusion we have shown that the unusual quantum Hall effect in graphene 
can be explained by
 the Hofstadter butterfly diagram for the honeycomb lattice.  
Because the honeycomb lattice  
has two sublattices in a unit cell, the similar situation for the half 
flux quantum per unit cell in the square lattice 
is realized in the small flux region.


\end{document}